\documentclass[aps,prb,reprint,superscriptaddress,showpacs]{revtex4-1}

\usepackage{amsmath}
\usepackage{graphicx}
\usepackage{natbib}

\begin{document}

\title{Perfect interference-less absorption at infrared frequencies by a van der Waal's crystal}

\author{D. G. Baranov}
\email[]{denis.baranov@phystech.edu}
\affiliation{Moscow Institute of Physics and Technology, 9 Institutskiy per., Dolgoprudny 141700, Russia}
\affiliation{All-Russia Research Institute of Automatics, 22 Sushchevskaya, Moscow 127055, Russia}

\author{J. H. Edgar}
\affiliation{Kansas State University, Department of Chemical Engineering, Durland Hall, Manhattan, KS 66506, USA}

\author{Tim Hoffman}
\affiliation{Kansas State University, Department of Chemical Engineering, Durland Hall, Manhattan, KS 66506, USA}

\author{Nabil Bassim}
\affiliation{U.S. Naval Research Laboratory, 4555 Overlook Ave, S.W., Washington, D.C. USA}

\author{Joshua D. Caldwell}
\affiliation{U.S. Naval Research Laboratory, 4555 Overlook Ave, S.W., Washington, D.C. USA}

\date{\today}

\begin{abstract}
Traditionally, efforts to achieve perfect absorption have required the use of complicated metamaterial-based structures as well as relying on destructive interference to eliminate back reflections. Here, we have demonstrated both theoretically and experimentally that such perfect absorption can be achieved using a naturally occurring material, hexagonal boron nitride (hBN) due to its high optical anisotropy without the requirement of interference effects to absorb the incident field. This effect was observed for p-polarized light within the mid-infrared spectral range, and we provide the full theory describing the origin of the perfect absorption as well as the methodology for achieving this effect with other materials. Furthermore, while this is reported for the uniaxial crystal hBN, this is equally applicable to biaxial crystals and more complicated crystal structures. Interference-less absorption is of fundamental interest to the field of optics; moreover, such materials may provide additional layers of flexibility in the design of frequency selective surfaces, absorbing coatings and sensing devices operating in the infrared.
\end{abstract}


\maketitle

\section{Introduction}
Perfect absorption of electromagnetic waves is at the cornerstone of a broad range of applied problems. In particular, strong electromagnetic absorption is highly desired in sensing and molecular detection \cite{Liu2010a, Kravets2013}, radar cloaking \cite{Petrov, Vinoy1996} and photovoltaics \cite{Luque2008}. Two different approaches towards complete absorption are typically used. The first incorporates engineered layered systems \cite{Kats2013,Shah1981}, amongst which the Dallenbach \cite{Dallenbach} and the Salisbury \cite{Shah1977} screens are the simplest examples. Alternatively, complete absorption can be achieved in metamaterial structures composed of resonant elements \cite{Landy2008} and periodic gratings \cite{Hutley,Popov}. The possibility of total absorption of light by a single nanoparticle has also been studied in a number of works \cite{Noh2012, Sentenac2013,Grigoriev2015}. In contrast to planar systems designed for absorption of plane waves, in this configuration total absorption requires coherent $4\pi$ illumination of a nanoparticle with a specially tailored electromagnetic field mode.

\begin{figure}[b]
\includegraphics[width=0.35\textwidth]{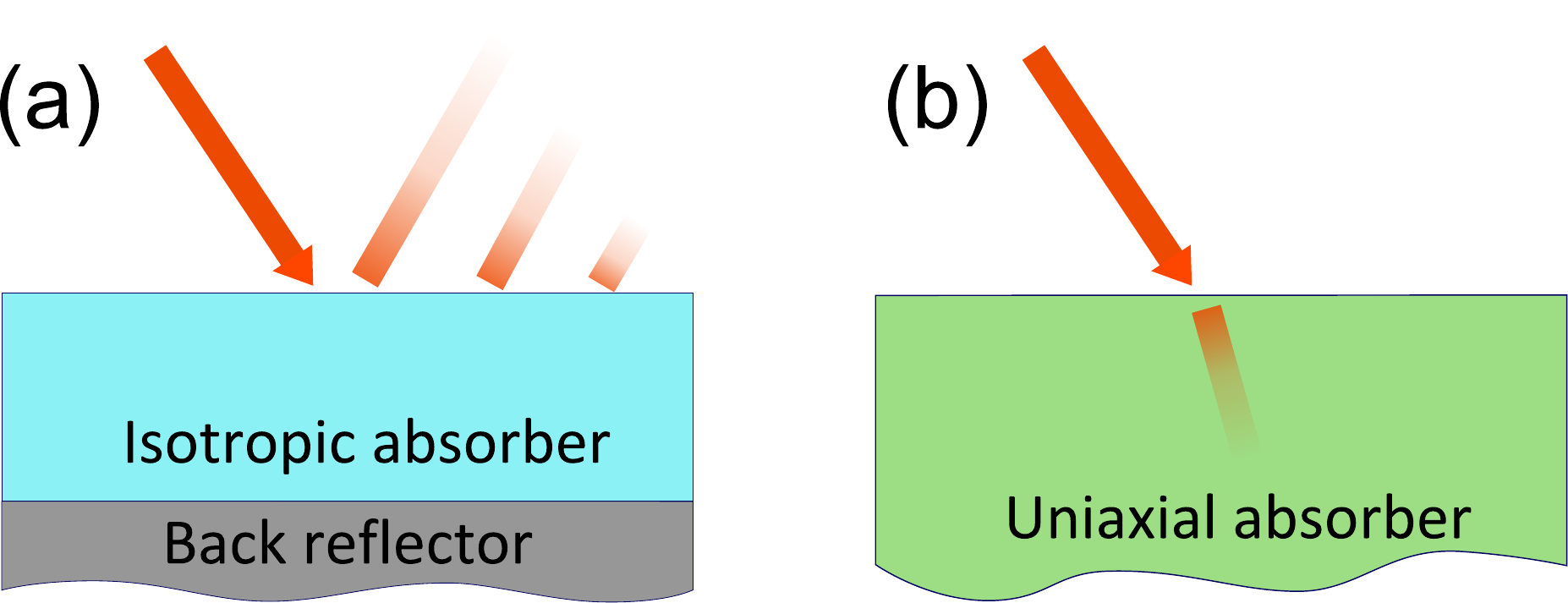}
\caption{\label{concept}Concept of an (a) interference-based and (b) interference-less perfect electromagnetic absorber.}
\end{figure}

In the light absorption schemes involving thin films and metamaterial structures, achieving zero reflection relies on the destructive interference of incident light, Fig.~\ref{concept}(a), and, thus, is highly sensitive to the geometry and thickness of the film \cite{Simovski15,Chen2012}. In recently developed two-port absorption devices, known as Coherent Perfect Absorbers (CPA) \cite{Chong2010, Wan2011, Zanotto2014}, absorption of light is also assisted by the destructive interference of two incoming plane waves. However, changes in the relative phases of the two beams violate the complete absorption, thereby removing the zero reflectance condition. In all such absorbers, the \emph{multi-pass} interference of light scattered from the structure is required for complete absorption.

In this Rapid Communication, we demonstrate experimentally that incident light can be completely absorbed in a \emph{one-pass} configuration that does not rely on interference effects, Fig.~\ref{concept}(b). To achieve this, the absorbing system consists of a half-space of a birefringent medium with low, but finite dielectric loss across the whole spectral range of interest. Under these conditions, at specific wavelengths and incidence angles, the amplitude of the reflected wave vanishes, leading to complete absorption of the incident wave inside the medium. Within the context of this work, we also verify these theoretical predictions experimentally through the use of a thick ($> 200$ $\mu$m) slab of the highly birefringent van der Waals crystal, hexagonal boron nitride (hBN) \cite{Caldwell2014}.
\section{Theory}
Let us consider the reflection of a p- (s-) polarized wave incident from air to a surface of an isotropic, non-magnetic medium. The reflection amplitude is given by \cite{Born1999}
\begin{equation} 
{r_p} = \frac{{\cos {\theta _i} - \frac{1}{{{\varepsilon }}}\sqrt {{\varepsilon} - {{\sin }^2}{\theta _i}} }}{{\cos {\theta _i} + \frac{1}{{{\varepsilon}}}\sqrt {{\varepsilon} - {{\sin }^2}{\theta _i}} }},
\label{eq1}
\end{equation}
\begin{equation}
{r_s} = \frac{{\cos {\theta _i} - \sqrt {{\varepsilon} - {{\sin }^2}{\theta _i}} }}{{\cos {\theta _i} + \sqrt {{\varepsilon} - {{\sin }^2}{\theta _i}} }},
\label{eq2}
\end{equation}
\begin{figure}
\includegraphics[width=0.4\textwidth]{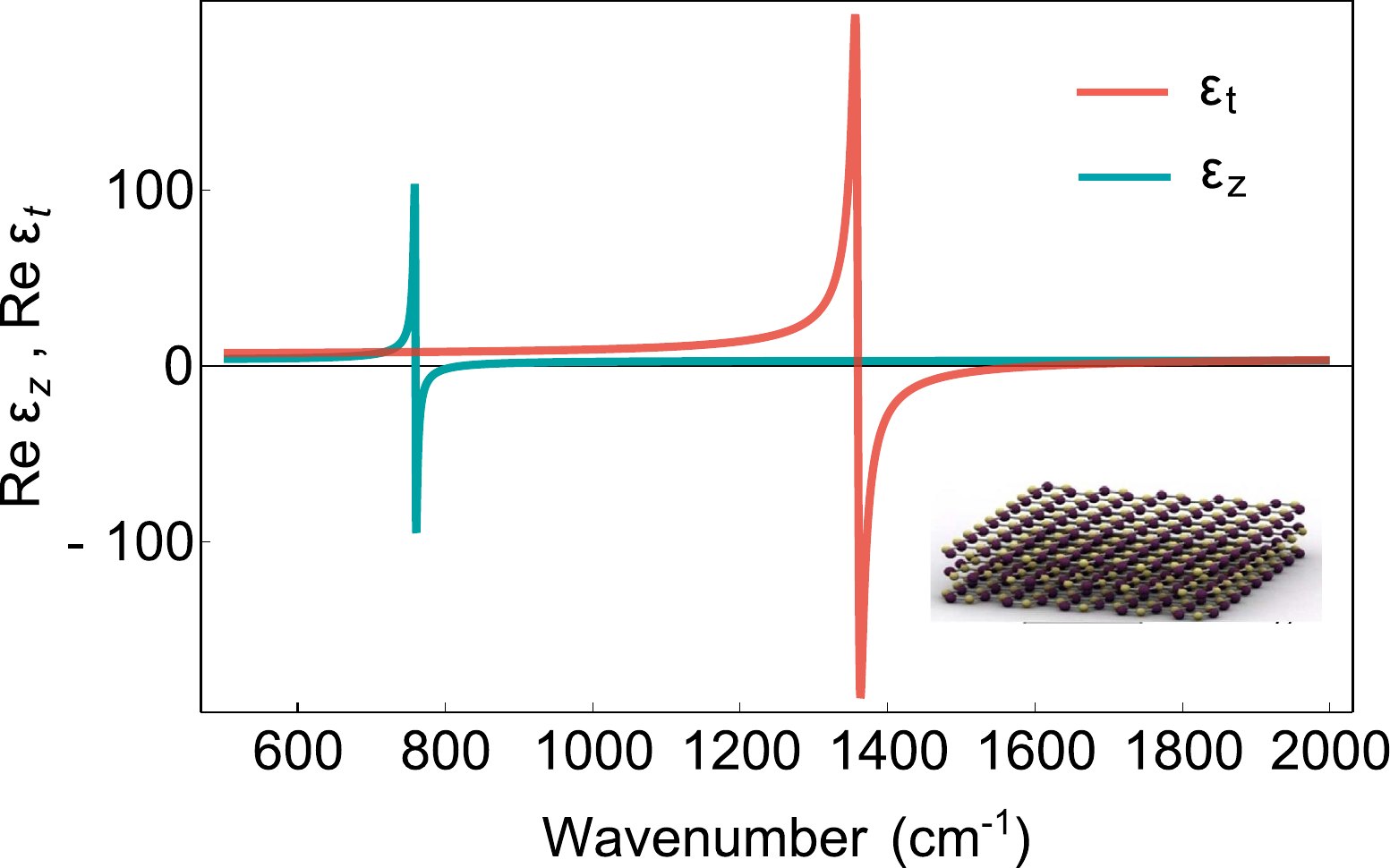}
\caption{\label{fig1} Frequency dispersion of the real parts of the in-plane ($\varepsilon _ t$) and out-of-plane ($\varepsilon _z$) hBN permittivity tensor components. Inset shows the schematic structure of the crystal.}
\end{figure}
where $\varepsilon$ is the complex permittivity of the medium and $\theta_i$ is the incident angle. 
In a lossless scenario, there exists a value of incidence angle, known as the Brewster angle, at which a p-polarized plane wave refracts into the dielectric medium without reflection. 
The presence of loss violates this perfect transmittance due to an impedance mismatch between air and the lossy medium, whose normalized impedance $Z=\frac{1}{{{\varepsilon}}}\sqrt {{\varepsilon} - {{\sin }^2}{\theta _i}}$ has a non-vanishing imaginary part associated with the electric loss in the dielectric.
Therefore, in order to absorb light within an isotropic medium, one places a reflector at the back surface of the lossy slab.
This allows for the cancellation of reflected light giving rise to a complete absorption at certain wavelengths and slab thicknesses despite a non-zero value of the one-pass reflection coefficient $r_{s,p}$.
Alternatively, one can put an anti-reflective coating \cite{ACRs} on top of the lossy medium, thereby resulting in a similar effect. Again, due to destructive interference, reflection may be suppressed at certain wavelengths allowing for complete absorption.

\begin{figure*}
\includegraphics[width=0.95\textwidth]{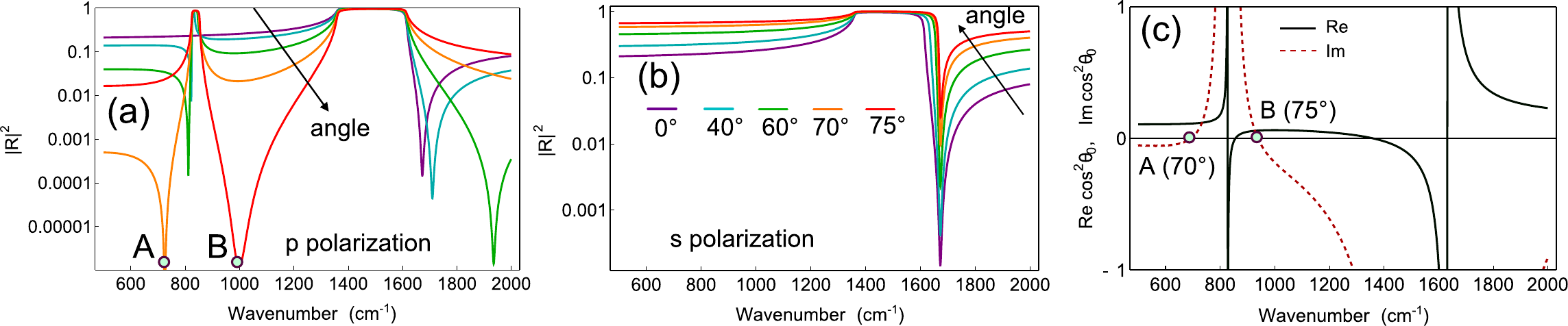}
\caption{\label{fig2} Reflection by a semi-infinite hBN crystal. (a, b) Calculated reflectance spectra for p and s polarizations, respectively, for a series of different incidence angles. (c) The real and imaginary parts of squared cosine of the prefect absorption angle $\theta_0$. At two points denoted by A and B on the graph, $\cos^2\theta_0$ becomes real and positive, corresponding to a real-valued incidence angle.}
\end{figure*}
The situation changes when the lossy medium is anisotropic. As we have theoretically shown earlier in Refs. \cite{Baranov2012, Baranov2012a}, a uniaxial bulk crystal with its optical axis normal to the interface can completely absorb incident p-polarized plane wave at a certain angle. The reflection coefficients for p and s polarization in this case read \cite{Lekner}
\begin{equation}
{\hat{r}_p} = \frac{{\cos \theta_i  - \sqrt {\left( {{\varepsilon _z } - {{\sin }^2}\theta_i } \right)/\left( {{\varepsilon _z }{\varepsilon _ t }} \right)} }}{{\cos \theta_i  + \sqrt {\left( {{\varepsilon _z } - {{\sin }^2}\theta_i } \right)/\left( {{\varepsilon _z }{\varepsilon _ t }} \right)} }},
\label{eq3}
\end{equation}
and
\begin{equation}
{\hat{r}_s} = \frac{{\cos \theta_i  - \sqrt {{\varepsilon _ t } - {{\sin }^2}\theta_i } }}{{\cos \theta_i  + \sqrt {{\varepsilon _ t } - {{\sin }^2}\theta_i } }},
\label{eq4}
\end{equation}
where $\varepsilon _t$ and $\varepsilon _z$ denote the permittivity components in directions normal and parallel to the optical axis of the birefringent crystal, respectively. From Eq. 3, one can derive the condition for zero reflection under p-polarized light irradiation as
\begin{equation}
({\varepsilon _z}{\varepsilon _ t } - 1){\cos ^2}\theta_0  = {\varepsilon _z } - 1,
\label{eq5}
\end{equation}
where $\theta_0$ denotes the incidence angle at which the incoming wave should impinge the surface for complete absorption. In contrast to the case of an isotropic material, this condition can be fulfilled even for a lossy material due to different values of permittivity components $\varepsilon_ t$ and $\varepsilon_z$. This behavior will also be prevalent for more complicated crystals structures, for instance those with biaxial anisotropy. We emphasize that, although scattering of light from uniaxial crystals, including absorbing media, was studied in a number of papers~\cite{Wooten68,Piro,Lekner}, the possibility of complete interference-less absorption in such crystals has not been predicted to the best of our knowledge. Note that for s polarization an incident wave feels only the in-plane permittivity component  $\varepsilon _ t$, thus the scattering of an incident wave occurs as if the material is isotropic. Correspondingly, vanishing reflection is met only in the trivial case $\varepsilon_t=1$. This is in accordance with the fact that the Brewster angle for s-polarized light can occur only in magnetic media \cite{Born1999}. 

Previously, such perfect absorption by an unbounded medium was investigated in artificial metamaterials consisting of parallel metallic nanowires embedded into a glass matrix \cite{Baranov2012}. However, the nature of such nanostructured materials brings specific limitations on the observation of certain effects predicted by the effective medium theory. In particular, the effective medium approach is no longer valid when the characteristic size $a$ of the metamaterial inclusion is comparable to the wavelength within the material \cite{Elser2007}. Moreover, as was shown in Ref. \cite{Belov2003}, strong spatial dispersion arises in the wire media even in the long wavelength regime. Following the recent discovery of hyperbolic behavior within the natural van der Waals crystal hBN by Dai et al. \cite{Dai2014} and Caldwell et al. \cite{Caldwell2014}, here we demonstrate that one-pass complete absorption also occurs within hBN, due to its natural crystal anisotropy within the mid-infrared spectral range.

As a two-dimensional van der Waal's crystal \cite{Geim2013}, hBN is highly anisotropic, consisting of planes of covalent, sp$^2-$bonded BN interlayed in a graphitic-like stack, with the layers bound only by van der Waals forces. Due to this crystal anisotropy, hBN is highly birefringent with different magnitudes for the in and out-of-plane dielectric function. Furthermore, as a polar crystal, it is capable of supporting surface phonon polariton modes in the spectral range between the longitudinal (LO) and transverse (TO) optic phonon frequencies with this regime referred to as the Reststrahlen band \cite{CaldwellNanoph}. For hBN, there are two sets of optic phonons corresponding to in- ($\omega_{LO,t}=1610$ cm$^{-1}$, $\omega_{TO,t}=1360$ cm$^{-1}$) and out-of-plane ($\omega_{LO,z}=825$ cm$^{-1}$, $\omega_{TO,z}=740$ cm$^{-1}$) vibrations referred to as the upper and lower Reststrahlen bands, respectively \cite{Dai2014,Caldwell2014}. Within these frequency bands, the coherent oscillations of the polar lattice result in a negative real part of the dielectric function. However, for hBN, the permittivity is negative along the transverse ($x,y$ plane) or optical ($z-$) axis of the crystal, within the upper and lower Reststrahlen bands, respectively, while the orthogonal component(s) of the dielectric function remain positive. This results in a hyperbolic behavior of the permittivity tensor, which can be clearly seen in Fig.~\ref{fig1} \cite{Poddubny}. Outside of these spectral bands, both the in- and out-of-plane components of the permittivity are positive, though highly birefringent. The hyperbolic nature of hBN has been demonstrated as a means to realize mid-infrared hyperlensing from a natural, unpatterned slab \cite{Taubner,Basov}, as well as allows one to increase lifetime of graphene plasmons and enables gate tunable hyperbolic response in graphene/hBN heterostructures \cite{Woessner,Caldwell2015,Kumar2015,Dai2015}.

Fig.~\ref{fig2}(a) shows the predicted reflectance spectra of p-polarized light incident on the surface of a semi-infinite hBN crystal for various values of the incidence angle. The calculations were performed using the dielectric function extracted by Caldwell et al. \cite{Caldwell2014} from exfoliated flakes of hBN. As discussed above, there are two high reflectivity Reststrahlen bands, which are clearly associated with the phonon-polariton resonances of hBN.
Note that the lower Reststrahlen band is absent for normally incident light (Fig.~\ref{fig2}(a), cyan curve). This occurs due to only $\varepsilon _z$ exhibiting resonant behavior at 820 cm$^{-1}$, while scattering of light is governed only by the in-plane permittivity $\varepsilon _t$ at normal incidence. 

Strong absorption under p-polarized excitation is demonstrated at different frequencies depending on the incidence angle, Fig.~\ref{fig2}(a). One can see a reflection dip occuring at 1650 cm$^{-1}$ at normal incidence that tends to higher frequencies with increasing incidence angle. A similarly strong absorption dip is observed near 800 cm$^{-1}$ and moves to the red with increasing angle. However, both vanish when the incidence angle exceeds $70 ^\circ$, but thus correlates with the development of a third high absorption dip that occurs between the two Reststrahlen bands.

The reflection spectra predicted for s-polarized incident light is quite different (Fig.~\ref{fig2}(b)), with only the higher frequency upper Reststrahlen band observed. In this case, only one high-frequency absorption dip occurs near 1700 cm$^{-1}$ that corresponds to the perfect transmittance condition for s-polarized light $\varepsilon_t=1$, and is therefore insensitive to the incident angle.

In order to examine the possibility of complete absorption (where one would have $\hat r_p$ being exactly zero), we plot in Fig.~\ref{fig2}(c) the real and imaginary parts of $\cos^2\theta_0$ deduced from Eq.~(\ref{eq5}). From this analysis, it follows that there are two points located at $\omega_A \approx 700$ cm$^{-1}$ (point A) and $\omega_B \approx 960$ cm$^{-1}$ (point B), where the imaginary part of $\cos^2\theta_0$ vanishes, while the real part is positive and bounded by the inequality $0<\cos^2\theta_0<1$ corresponding to a real-valued incidence angle. Therefore, at these two conditions a p-polarized plane wave incident on a half-space of hBN would be refracted into the lossy medium \emph{without reflection} and therefore completely absorbed inside.

\section{Experiment} 
In order to experimentally confirm our theoretical predictions, we grew hBN crystals via the Ni-Cr flux growth process, based on methods described by Hoffman et al. \cite{Hoffman} and Kubota et al. \cite{Kubota}. In this process, fine grain hBN powder was dissolved into a molten Ni-Cr flux (51 wt $\%$ Cr basis) in a resistively-heated furnace under nitrogen atmosphere at high temperature ($1500 ^\circ$ C). Then the flux was driven into supersaturation by slow cooling ($4^\circ$ C/h), causing the hBN crystals to nucleate and grow on the metal surface. The crystals analyzed were optically clear, with a pillar-like structure and an apparent surface grain size of 250-500 $\mu$m and a thickness of approximately 200 $\mu$m revealed by Focused Ion Beam-SEM cross-sectioning.

A large flat area ($>2$ cm) enabled the measurement of FTIR reflectance as a function of angle between near-normal and close to grazing incidence, with both s- and p-polarized light using the Seagull attachment for a Thermoscientific FTIR spectrometer. The spectra were collected in reference to a flat aluminum mirror with a spectral resolution of 2 cm$^{-1}$.  

\begin{figure}
\includegraphics[width=0.5\textwidth]{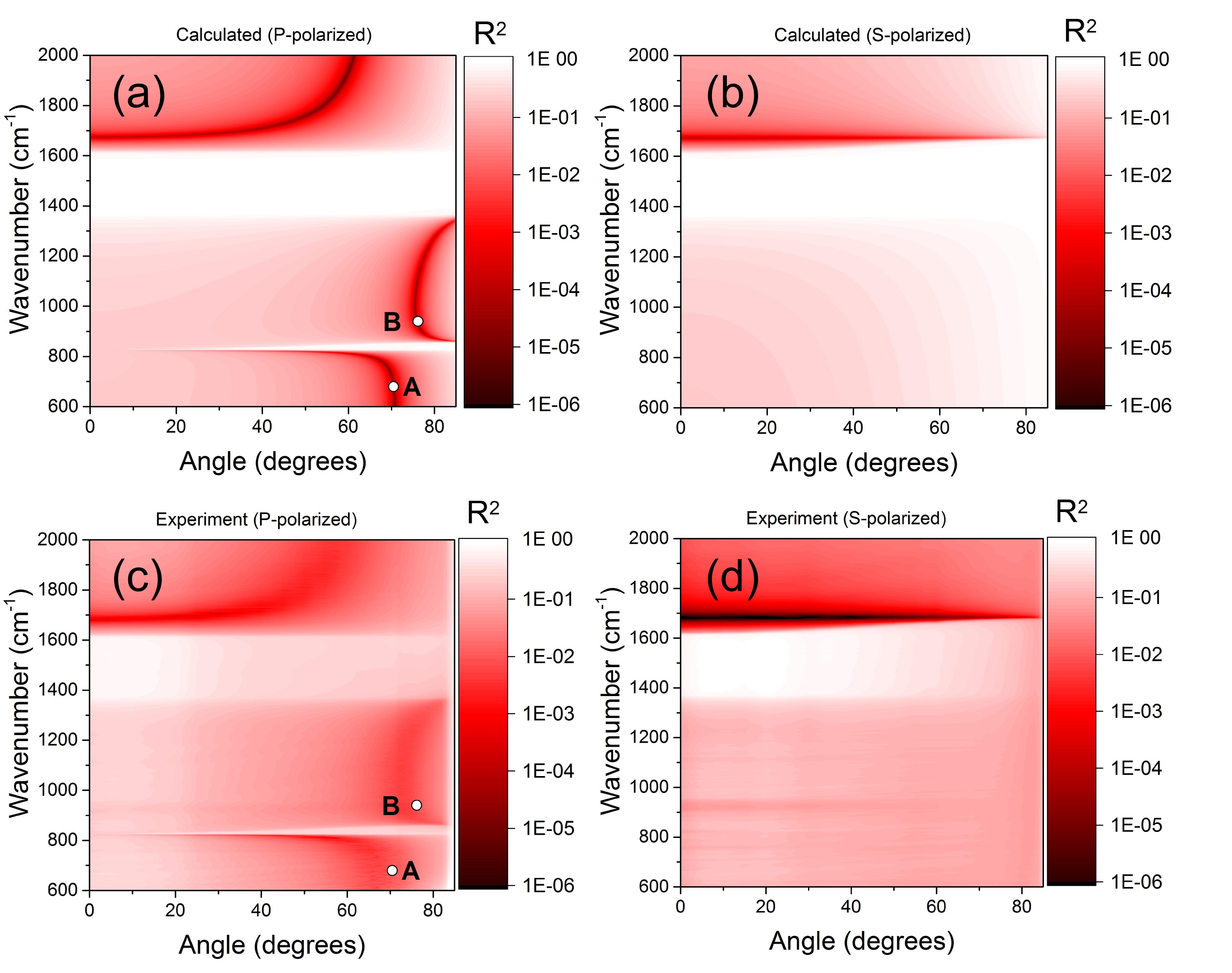}
\caption{\label{fig3}(a,b) The calculated reflectivity spectra of an hBN crystal for p and s polarization of incident light, respectively. (c,d) The corresponding measured spectra for 200 $\mu$m thick hBN slab on a metal substrate. The points A and B mark the position of perfect absorption points corresponding to those in Fig. 2(c)}
\end{figure}

In order to best visualize the changes in reflectivity for both p- and s-polarizations as a function of angle, density plots are provided in Figs.~\ref{fig3}.
Data for s-polarized incidence at angles larger than $85 ^\circ$ were not plotted due to potential measurement artifacts.
Overall, the theoretical predictions ('Calculated'; Figs.~\ref{fig3}(a),(b)) and experimental data ('Experiment'; Figs.~\ref{fig3}(c),(d)) are in a good agreement. Due to presence on an underlying metal substrate, absorbance by the hBN crystal is related to reflectance via $A=1-R^2$.
The three different regions of attenuated reflection (dark regions in Figs.~\ref{fig3}) for p-polarized irradiation and one for s-polarized light are observed and their positions are in excellent quantitative agreement with the theoretical predictions.
As mentioned above, only positions marked by 'A' and 'B' correspond to the true perfect absorption conditions, while the reflection dip for s-polarized light originates from the perfect transmission condition, thus the absorption cannot be perfect for this polarization.

\begin{figure}
\includegraphics[width=0.5\textwidth]{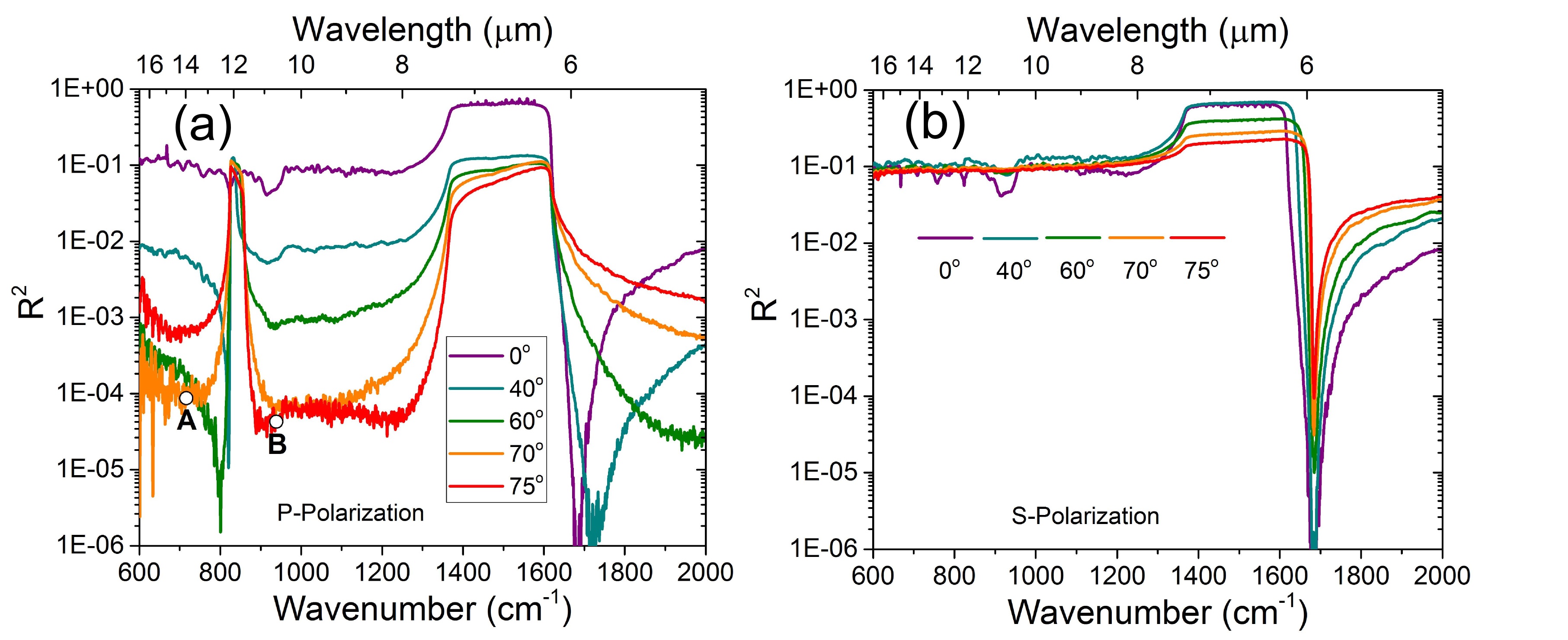}
\caption{\label{fig4}Experimental reflectivity spectra for a series of incidence angles for (a) p and (b) s polarizations. The points A and B mark the position of perfect absorption points.}
\end{figure}

To explore the capabilities of hBN as an interference-less electromagnetic perfect absorber, let us turn to one-dimensional experimental reflectivity spectra for a series of incidence angles, consistent with those presented in Figs.~\ref{fig2}(a) and (b), depicted in Figs.~\ref{fig4}. At the frequencies of expected perfect absorption for p-polarized light $\omega_A$ and $\omega_B$ the reflection amplitudes are as low as $10^{-4}$ and $3 \cdot 10^{-5}$, respectively, Fig.~\ref{fig4}(a), what can be regarded as almost complete absorption for most practical applications. 

At 800 cm$^{-1}$ and 1700 cm$^{-1}$, where complete absorption is not predicted by Eq.~(\ref{eq5}), reflection reaches even lower values on the order of $10^{-6}$. Thus, while these latter two positions do not correspond to perfect absorption, in terms of applications where reflectivity is the mode of operation, this would amount to a similar response. Larger observed absorption at these two frequencies may be attributed to relatively low angle resolution of our setup. There is also a noticeable difference between the experimental and theoretical data within the upper Reststrahlen band. Measurements demonstrate decrease of the reflectivity upon increasing incidence angle, while Fresnel formulas predict near unity reflectance for a wide range of angles, Fig.~\ref{fig2}(a,b). This discrepancy possibly originates from the surface roughness of the hBN sample, a wide range of crystal orientations in the polycrystalline sample (c-axis alignment is always preserved, but rotation of crystals in plane is of course present) and the following excitation of high$-k$ bulk modes allowed within the Reststrahlen band which are strongly absorbed inside the crystal.

Notably, strongly attenuated reflection can be observed for s-polarized light as well, Fig.~\ref{fig4}(b). The reflection reaches value $10^{-6}$ at 1700 cm$^{-1}$ at normal incidence. 
The spectral position of this reflectivity dip is almost unaffected by the incidence angle value with the magnitude of the reflectivity monotonically increasing from $10^{-6}$ up to $10^{-4}$ at $85 ^\circ$.
This reflection dip is related to the condition of perfect transmittance for s polarization $\varepsilon _ t=1$, as discussed above. 
Such high tolerance to the incidence angle variation makes sufficiently thick layer of hBN an omnidirectional absorber at 1700 cm$^{-1}$.
This property can be understood from Eq.~(\ref{eq4}); indeed, for the in-plane permittivity ${\varepsilon _t} = 1 + i\varepsilon ''_t,~\varepsilon ''_t \ll 1$ the reflection amplitude (\ref{eq4}) can be expanded into a Taylor series:
\begin{equation}
{\hat r_s} \approx - \frac{{i{\varepsilon ''_t}}}{{4{{\cos }^2}{\theta _i}}}.
\label{eq6}
\end{equation}
Therefore, at any incidence angle, the incoming wave is almost completely transmitted into the underlying absorbing material, whereas only a small fraction of the energy proportional to ${\left| {{\varepsilon ''_t}} \right|^2}$ undergoes reflection back into air.

\section{Concluding remarks}
In conclusion, we have demonstrated that incident light can be completely absorbed in a one-pass configuration, which does not rely on interference of scattered plane waves. The mechanism of perfect absorption is the impedance matching between air (vacuum) and the birefringent absorbing semi-infinite medium. The concept of one-pass perfect absorption was verified experimentally at IR frequencies with a thick slab of the birefringent van der Waal's crystal hexagonal boron nitride. We measured the angular dependence of the reflectivity spectra for both p and s polarization and observed pronounced reflectivity dips whose position agrees with the theoretical predictions.

The concepts laid out here provide a unique method for realizing perfect absorption of electromagnetic fields without the incorporation of interference effects. Such an approach is of specific value where the phase of the field must be maintained and where simplified, semiconductor compatible devices are desired or required. 
While such an approach using uniaxial crystals may not provide exclusive benefits for device miniaturization due to the requirement that the absorbing slab be comparable in thickness to the skin depth of the radiation within that medium, the use of biaxial crystals with anisotropy along all three crystallographic directions could provide unique opportunities for spreading the absorption into a full three dimensional volume. 
In addition, the reported theory and experimental observations are of fundamental interest to the field of optics, since they show the possibility of complete absorption of light without destructive interference. 
Therefore, this effect offers a new tool for controlling electromagnetic absorption and may have implications for the design of sensing and photovoltaic devices. 
Moreover, knowing the extraordinary optical properties of hBN will further stimulate the development of novel electronic devices implementing various van der Waal's crystals, all of which should offer high birefringence and varying skin depths due to the highly anisotropic van der Waals crystal structure and variation in electronic and optical properties.

\begin{acknowledgments}
D.G.B. acknowledges funding provided by RFBR project No 13-07-92660 and Dynasty Foundation. Funding for J.D.C. and N.D.B. was provided by the Office of Naval Research and administered by the Nanoscience Institute of the Naval Research Laboratory in Washington, D.C. J.H.E and T.H. greatly appreciate the support of the Department of Homeland Security and the II-VI Foundation. NRL coauthors would like to express their thanks to Dr. Blake Simpkins for assistance with the angle-resolved FTIR experimental setup. D.G.B. would like to thank A. P. Vinogradov and C. R. Simovski for stimulating discussion.
\end{acknowledgments}

\bibliography{absorbers}

\end{document}